# Superconductivity in Fe and As based compounds: a bridge between $MgB_2$ and cuprates

A. Bussmann-Holder[1], A. Simon[1], H. Keller[2] and A. R. Bishop[3]

[1]Max-Planck-Institut für Festkörperforschung, Heisenbergstr. 1, D-70569 Stuttgart, Germany

[2]Physik Institut der Universität Zürich, Winterthurerstr. 190, CH-8057 Zürich, Switzerland

[3]Los Alamos National Laboratory, Los Alamos, NM87545, USA

By interpreting various experimental data for the new high temperature FeAs type superconductors in terms of lattice mediated multi gap superconductivity, it is shown that these systems strongly resemble $MgB_2$, however, with the distinction that *local* polaronic lattice effects exist. This fact establishes a connection to cuprate high temperature superconductors where polaron formation is essential for the pseudogap phase and the unconventional isotope effects observed there. However, similar to $MgB_2$ and in contrast to cuprates the two superconducting gaps in the Fe-As based materials are isotropic s-wave gaps.

The discovery of high temperature superconductivity in Fe-As type compounds [1, 2] has produced a surge in solid state research since their maximum $T_c$ value well exceeds that in $MgB_2$ and places them in the vicinity of cuprate superconductors. Their proximity to a magnetic phase has invoked the possibility that strong correlations dominate the superconductivity [3], and that electron-lattice interactions are too weak to account for the pairing mechanism and the high values of $T_c$ [4]. In addition, first principles calculations of the electron-phonon coupling constant [4] together with inelastic neutron scattering studies [5, 6] showed that the coupling is small and cannot directly cause high temperature superconductivity. On the other hand, local structural probes such as extended x-ray absorption fine structure (EXAFS) measurements reveal an anomalous temperature dependence of local Fe-As lattice displacements [7] which closely resembles the ones observed in cuprate superconductors [8]. This observation suggests that *local* rather than *global* electron-lattice interaction effects are involved and that polaron formation is responsible for the anomalies observed in EXAFS data [7].

In addition to the local lattice instabilities, the complex Fermi surface topology points to multiband superconductivity [9, 10] analogous to $MgB_2$. Tunneling experiments [11], infrared spectroscopy [12], specific heat [13], lower critical field measurements [14], point contact spectroscopy [15], as well as penetration depth results [16 – 21] and magnetic torque studies [22] show considerable deviations from single gap behavior and have been interpreted in terms of multiple isotropic gaps.

In the following we use a multiband model with polaronic couplings [23] to quantitatively demonstrate that Fe-As type superconductors are two-gap s-wave superconductors with strong local electron lattice coupling. While the former property places these systems in the company of $MgB_2$, the latter links them more to cuprates where polaron formation onsets at high temperatures [24]. However, opposite to cuprates, polaron coherence does not take place, i.e., a stripe like phase with preformed pairs is absent, consequently the $T_c$'s of these compounds will probably not reach the values of cuprates. This seems to be confirmed experimentally since increasing the chemical pressure does not lead to further enhancements of $T_c$ beyond a certain critical pressure but rather a reduction [25]. We start from an effective BCS-type Hamiltonian extended to account for two-gap superconductivity with interband coupling [26]:

$$H = H_0 + H_1 + H_2 + H_{12}$$

$$H_0 = \sum_{k_1\sigma} \xi_{k_1} c^+_{k_1\sigma} c_{k_1\sigma} + \sum_{k_2\sigma} \xi_{k_2} d^+_{k_2\sigma} d_{k_2\sigma}$$

$$H_1 = \sum_{k_1 k'_1 q} V_1(k_1,k'_1) c^+_{k_1+q/2\uparrow} c^+_{-k_1+q/2\downarrow} c_{-k'_1+q/2\downarrow} c_{k'_1+q/2\uparrow}$$

$$H_2 = \sum_{k_2 k'_2 q} V_2(k_2,k'_2) d^+_{k_2+q/2\uparrow} d^+_{-k_2+q/2\downarrow} d_{-k'_2+q/2\downarrow} d_{k'_2+q/2\uparrow}$$

$$H_{12} = \sum_{k_1 k_2 q} V_{12}(k_1,k_2) \{ c^+_{k_1+q/2\uparrow} c^+_{-k_1+q/2\downarrow} d_{-k_2+q/2\downarrow} d_{k_2+q/2\uparrow} + h.c. \} . \tag{1}$$

Here, $c^+$, $c$, $d^+$, $d$ are (electron / hole) creation and annihilation operators of the As p-Fe d hybridized bands in the FeAs sheets with momentum k dependent energies $\xi$. The intraband pairing potentials $V_i$ are governed by electron lattice interactions, which have a sizable strength in one channel only, whereas they are too small in the other channel to support superconductivity alone. The interband interaction $V_{12}$ is explicitly assumed to be multiphonon mediated, stemming from strong anharmonicity, as observed experimentally [7, 27]. Including local polaronic couplings means that we modify $H_0$ as:

$$\overline{H}_0 = \sum_{k_i\sigma} \xi_{k_i} c^+_{k_i\sigma} c_{k_i\sigma} + \sum_q \hbar\omega_q b^+_q b_q + \frac{1}{\sqrt{N}} \sum_{q,\sigma,k_i} \gamma_i(q) c^+_{k_i+q\sigma} c_{k_i\sigma} (b_q + b^+_{-q}); i = 1,2 , \tag{2}$$

where $\hbar\omega$ is the momentum $q$ dependent phonon energy, $b$, $b^+$ are phonon annihilation and creation operators, and $\gamma$ is the local electron-lattice coupling. By using the canonical Lang-Firsov transformation [28], electron and lattice degrees of freedom can be decoupled, whereby a rigid harmonic oscillator shift of the phonon operators takes place while electronic energies are exponentially reduced to $\tilde{\xi}_k = \xi_k \exp[-\gamma^2 \coth(\hbar\omega/2kT)]$. The superfluid stiffness is calculated within linear response theory through the relation between the current and the induced transverse gauge field [29]:

$$\rho_s^\alpha = \frac{1}{2N} \sum_k \left\{ \left( \frac{\partial \tilde{\xi}_{k_i}}{\partial k_\alpha} \right)^2 \frac{\partial f(E_{k_i})}{\partial E_{k_i}} + \frac{1}{2} \frac{\partial^2 \tilde{\xi}_{k_i}}{\partial k_i^2} \left[ 1 - \frac{\tilde{\xi}_{k_i}}{E_{k_i}} \tanh \frac{E_{k_i}}{2kT} \right] \right\}. \tag{3}$$

Here $E_k = \sqrt{\tilde{\xi}^2 + \overline{\Delta}_i^2}$, where the values of $\overline{\Delta}_i$ are obtained from Eq. 1 by applying standard techniques:

$$< c^+_{k_1\uparrow} c^+_{-k_1\downarrow} > = \frac{\overline{\Delta}_{k_1}}{2E_{k_1}} \tanh \frac{\beta E_{k_1}}{2} = \overline{\Delta}_{k_1} \Phi_{k_1}$$

$$< d^+_{k_2\uparrow} d^+_{-k_2\downarrow} > = \frac{\overline{\Delta}_{k_2}}{2E_{k_2}} \tanh \frac{\beta E_{k_2}}{2} = \overline{\Delta}_{k_2} \Phi_{k_2}$$

$$\overline{\Delta}_{k_2} = \sum_{k_2'} V_2(k_2, k_2') \overline{\Delta}_{k_2'} \Phi_{k_2'} + \sum_{k_1} V_{1,2}(k_1, k_2) \overline{\Delta}_{k_1} \Phi_{k_1}$$

$$\overline{\Delta}_{k_2} = \sum_{k_2'} V_2(k_2, k_2') \overline{\Delta}_{k_2'} \Phi_{k_2'} + \sum_{k_1} V_{2,1}(k_2, k_1) \overline{\Delta}_{k_1} \Phi_{k_1}. \tag{4}$$

The above coupled gap equations have to be solved simultaneously and self-consistently in order to derive their temperature dependence and $T_c$. Guided by experimental data [16 – 19], both gaps are taken to be isotropic s-wave. In Figs. 1 we compare the theoretically derived results for the normalized in-plane penetration depth $\lambda_{ab}^{-2}(T)/\lambda_{ab}^{-2}(0) = \sigma_{sc}(T)$ to experimental data for two Fe type superconductors, namely, Ba$_{0.6}$K$_{0.4}$Fe$_2$As$_2$ [16], Ba$_{1-x}$K$_x$Fe$_2$As$_2$ [19]. (Note, that x is undefined in the latter compound). The good agreement between experiment and theory supports the present approach. Also shown in the figures are the individual contributions $\omega\rho_{sc}^s$ and $(1-\omega)\rho_{sc}^l$ to the total superfluid densities: $\rho_{sc}(T) = \lambda_{ab}^{-2}(T)/\lambda_{ab}^{-2}(0). = \omega\rho_{sc}^s + (1-\omega)\rho_{sc}^l$, where $\rho_{sc}^s$ and $\rho_{sc}^l$ are the individual normalized s-wave superconducting densities of the small, large gap, respectively, and $\omega$ is the weighting factor. The ratios $\omega/(1-\omega)$ from the small $\rho_{sc}^s$ and the large density $\rho_{sc}^l$ contributions vary considerably with the transition temperature: While for the smaller $T_c$ system, Ba$_{1-x}$K$_x$Fe$_2$As$_2$, $\rho_{sc}^l$ dominates the temperature dependence of the penetration depth, this is reversed for the higher $T_c$ compound, Ba$_{1-x}$K$_x$Fe$_2$As$_2$, where the major contribution stems from $\rho_{sc}^s$. In both cases, however, an inflection point at low temperatures is present which demonstrates that the two contributing gaps must have substantially different gap values, i.e., a small one coexists with a large one where their difference in size is a factor of three or more. This is in full agreement with experimental findings [16 – 21]. In Fig. 1c we show the variations of the temperature dependence of the normalized penetration depth versus $t = T/T_c$ for various ratios of the two superfluid density contributions. Obviously, the temperature dependence varies strongly with this ratio and may thus explain the non-exponential behavior at low temperatures of the superfluid density observed experimentally [20].

From the present analysis it appears that for small $T_c$ an almost single superfluid density dominates the penetration depth. With increasing $T_c$ the contributions from the individual superfluid densities approach each other, to be about equally contributing at optimal values of $T_c$ thus achieving the maximum enhancement mechanism of the two-band model. Beyond this value the trend reverses and the leading gap changes its character, analogous to Al doped MgB$_2$ [30], to approach again the nearly single superfluid density behavior. In order to demonstrate this evolution, we have calculated the penetration depth for values of $T_c$ smaller than and intermediate between those displayed in Figs. 1 and near optimal doping where $T_c=50K$. Experimental data for this regime are currently absent.

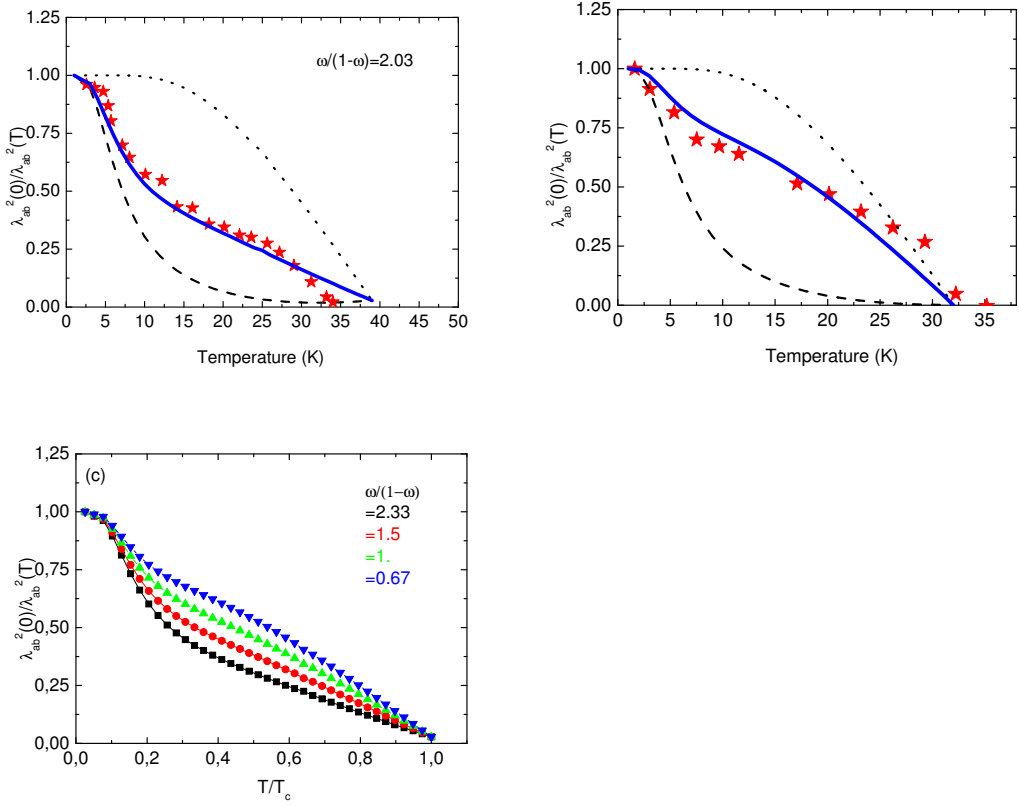

**Figure 1 (a, b)** Comparison of experimental (full stars) and theoretical (full lines) temperature dependencies of the normalized in-plane penetration depths together with the individual contributions from the two components $\rho_{sc}^s$ (dashed lines) and $\rho_{sc}^l$ (dotted lines). In (a) the experimental data have been taken from Ref. 16 for $Ba_{0.6}K_{0.4}Fe_2As_2$. In (b) the data from Ref. 19 for $Ba_{1-x}K_xFe_2As_2$ have been plotted. The model parameters for both systems are given in Table 1 together with theoretically derived ones shown in Fig. 2. **(c)** Calculated temperature dependencies for varying $\omega/(1-\omega)$ to the total $\rho_{sc}(T) = \lambda_{ab}^{-2}(T)/\lambda_{ab}^{-2}(0). = \omega\rho_{sc}^s + (1-\omega)\rho_{sc}^l$.

If the above conclusions regarding the individual superfluid density contributions to the penetration depth are correct then we would expect that the small to large superfluid density ratio $\omega/(1-\omega)$ crucially influences the value of $T_c$. This trend is shown in Fig. 2, which can be tested experimentally by systematically measuring the superfluid density as a function of doping. Note, that the maximum value of $T_c$ is predicted to be 56K for $\omega/(1-\omega) = 1.55$.

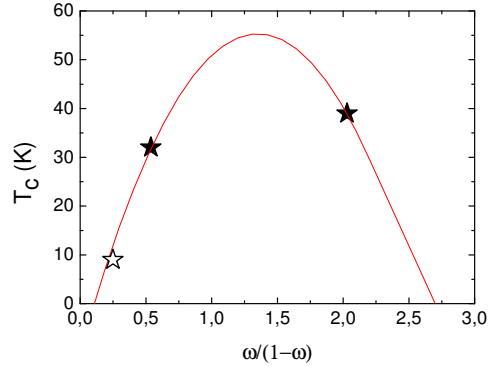

**Figure 2** Calculated dependence of $T_c$ on the ratio $\omega/(1-\omega)$. The full line is our prediction of how $T_c$ should depend on this ratio. The full stars refer to the values used in Figs. 1a, b and as derived for $FeSe_x$ (open star) from Ref. 21. For all three systems the parameters used in the calculation are given in Table 1.

In order to justify our polaronic approach (Eq. 2), we investigate the effect of polaron formation on the lattice. As already mentioned above, the phonon energies undergo a rigid oscillator shift due to polaron formation, where this shift is proportional to the coupling constant $\gamma$ and the band energies, eq 2. This provides a substantial temperature dependent renormalization of the local phonon mode frequency which in turn leads to a pronounced effect on the mean square Fe-As displacement. Since the renormalized frequency experiences pronounced softening, the related displacement inversely increases and is maximal when the softening is complete [24]. These effects have been observed in cuprate superconductors and have been related to the onset of the stripe phase [24]. Actually, in cuprates not only one divergence in the mean-square Cu-O displacement has been observed but two: one at the onset temperature T* of the pseudogap phase, and another one either just at or slightly below $T_c$ depending on which displacement is tested [see Refs. 8, 24 and references therein]. These results have been interpreted as signatures of polaron coherence at T* and persistence into the superconducting phase, T<$T_c$. The EXAFS experiments on the Fe-As type superconductors [7] show, in contrast to $MgB_2$ where no such effects are present, a similar divergence in the mean square Fe-As displacement as seen with Cu-O, however with only one anomaly appearing just at $T_c$. This observation suggests that polaron formation is effective also in pnictides but only gains coherence at $T_c$ and not at a higher pseudogap temperature. The weaker electron-lattice coupling in Fe-As based compounds as compared to cuprates enhances the polaron mobility and screens the polaron-polaron interaction which – in turn – suppresses stripe formation.

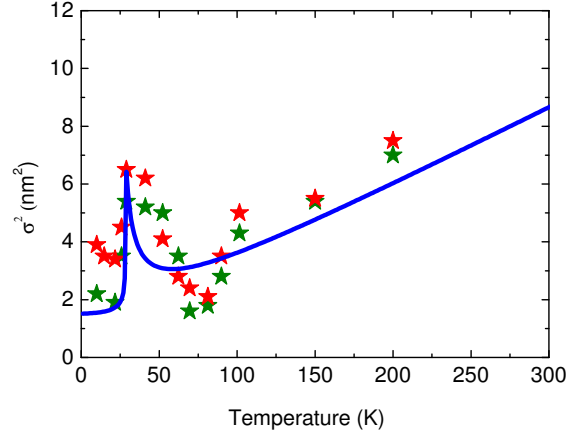

**Figure 3** Temperature dependence of the mean square Fe-As displacement. Full stars correspond to experimental data from Ref. 7 for LaFeAsO$_{0.93}$F$_{0.07}$; the full line is theoretically derived using the procedure described in Ref. 24.

The comparison between the EXAFS data [7] and the numerical results is shown in Fig. 3. Note, that in order to avoid the use of too many parameters, no damping in the theoretical curve is included. In spite of this restriction, good agreement between theory and experiment is achieved. The polaronic coupling which has been used, is substantially smaller than that obtained at optimum doping in cuprates ($\gamma = 0.75 - 0.3$ in cuprates [24], $\gamma = 0.2$ in Fe based compounds), indicating that lattice effects in the Fe based systems are less prevalent than in cuprates. However, in order to obtain the divergence observed experimentally, polaron formation has to play a significant role also in these superconductors. This should be noticeable in an isotope effect on the penetration depth which is predicted to be small due to the rather small (of order a few %) value of $\gamma$.

**Table 1 Captions:** Parameters used to calculate the results shown in Figs. 1 a, b, and those of Ref. 21, not shown. The first column refers to the superconducting transition temperature T$_c$, the second column is the intraband interaction V multiplied by the density of states at the Fermi energy in the small gap, the third is the same but for the large gap, the fourth column is the interband interaction between both bands multiplied by the interband density of states at the Fermi energy*. The energies for the small and the large superconducting gaps (exp. values are given in parenthesis) and the corresponding ratios of the superfluid density contributions used in the calculations are displayed in columns 5 to 7.

**Table 1**

| T$_c$ (K) | V$_l$N$_l$(0) | V$_s$N$_s$(0) | V$_{sl}$N$_{sl}$(0) | $\Delta_s$(meV)* | $\Delta_l$ (meV) | $\omega/(1-\omega)$ |
|---|---|---|---|---|---|---|
| 9 | 0.208 | 0.018 | 0.06 | 0.71 | 2.42 | 0.25 |
|  |  |  |  | 0.38(1)[a] | 1.6(2)[a] |  |
| 32 | 0.297 | 0.018 | 0.06 | 1.71 | 8.32 | 0.54 |

| | | | | 1.5(2)[b] | 9.1(2)[b] | |
|---|---|---|---|---|---|---|
| 39 | 0.317 | 0.018 | 0.06 | 1.95 | 10.09 | 2.03 |
| | | | | 2.0(3)[c] | 8.9(4)[c] | |

[a]Ref. 21, [b]Ref. 16, [c]Ref. 19

*Note that $N_sV_s(0)$ and $N_{sl}V_{sl}(0)$ are assumed to be constant for all values of $T_c$ in order to minimize the number of parameters. In spite of this simplification, $\Delta_s$ increases systematically with increasing $T_c$ emphasizing the gap coupling effect through the interband channel. The difference between calculated and experimental gap values arises from the fact that the gap coupling is not included in the experimentally derived values.

In conclusion, we have shown that Fe-based superconductors exhibit features common to both $MgB_2$ and cuprates. The link is the existence of multiple gaps, with the distinction from cuprates [31] but in analogy to $MgB_2$, of all being isotropic s-wave gaps. In contrast to $MgB_2$ *local* lattice effects play an important role for superconductivity, however being less effective than in cuprates. It thus seems, that pnictide superconductors lie in between these two materials and consequently have limited values of $T_c$ which will not reach the high values observed in cuprates. It should be pointed out, however, that the multiband scenario used here and supported by many experiments [8 − 22] is questioned in Ref. 32, where a single d-wave gap has been reported. Similarly, the role of the lattice for superconductivity in Fe based compounds has been questioned in Refs. 3 − 6. Recent isotope experiments [33], on the other hand, support this approach. Since we make specific predictions, our model can be tested experimentally.

**Acknowledgement** It is a pleasure to acknowledge stimulating discussions with and a critical reading of the manuscript by K. A. Müller.